# Title: Gate-tunable chiral spin mode in $WSe_2/WS_2$ moiré superlattices

**Authors:** Zhexu Shan[1,2]†, Wenjian Su[1,2]†, Kenji Watanabe[3], Takashi Taniguchi[4], Shiyao Zhu[1,2,5,6,7], Yuanfeng Xu[1,8]* and Yanhao Tang[1,2]*

**Affiliations:**
[1] School of Physics, Zhejiang University, Hangzhou 310027, China
[2] Zhejiang Key Laboratory of Micro-Nano Quantum Chips and Quantum Control, Zhejiang University, Hangzhou 310027, China
[3] Research Center for Electronic and Optical Materials, National Institute for Materials Science, 1-1 Namiki, Tsukuba 305-0044, Japan
[4] Research Center for Materials Nanoarchitectonics, National Institute for Materials Science, 1-1 Namiki, Tsukuba 305-0044, Japan
[5] State Key Laboratory for Extreme Photonics and Instrumentation, Zhejiang University, Hangzhou 310027, China
[6] College of Optical Science and Engineering, Zhejiang University, Hangzhou 310027, China
[7] Hefei National Laboratory, Hefei 230088, China
[8] Center for Correlated Matter, Zhejiang University, Hangzhou 310027, China
*Corresponding author. Email: y.xu@zju.edu.cn, yanhaotc@zju.edu.cn
† These authors contributed equally

**Abstract:**
The interplay between many-body and spin-orbit effects (SOC) can lead to novel phenomena in solids. Chiral spin modes (CSM) are collective spin excitations that arise from such interplay and connect states with opposite chirality, which, however, have been rarely observed. Here, we report a gate-tunable CSM that occurs between conduction SOC-split minibands in near-0°-twisted $WSe_2/WS_2$ bilayers. This mode manifests as a sharp resonance with giant Raman efficiency in the pseudovector-symmetry channel of the Raman spectra. By varying fillings, the CSM transitions from chiral spin exciton to excitonic polaron. The spin-flip nature is directly confirmed by the Zeeman effect. Moreover, the filling dependence compellingly evidences a charge-transfer insulator at filling of one. Our results demonstrate transition-metal-dichalcogenides moiré superlattices as a fertile platform for exploring exotic low-lying collective excitations.

**Main Text:**
The many-body and SOC interactions together are key ingredients of exotic states in solids, such as unconventional magnetism and correlated topological phases(*1*). The chiral spin mode(*2*, *3*) (CSM)—an exotic type of collective spin excitations, is another example, in which transitions occur between SOC-split states with opposite chirality and many-body interactions lead to collective effects. In contrast with conventional collective spin excitations that widely exist in magnetic materials(*4*), the CSM can exist in non-magnetic systems with inversion symmetry broken, where the SOC plays a role of an effective magnetic field and time-reversal symmetry remains intact. These modes have been theoretically proposed in general two-dimensional Fermi liquids with the SOC(*2, 3*). However, the experimental observations remain scarce, only limited to two materials—$Bi_2Se_3$ topological insulator(*5*) and Mn-doped CdTe quantum wells(*6*).

The semiconductor transition-metal-dichalcogenides (TMD) moiré superlattices(*7*) provide intriguing platforms for realizing the CSM, owing to strong electron correlation arising from flat minibands(*8*) and large Ising SOC inherent to TMD monolayers(*9*), as well as the great tunability. The TMD moiré superlattices have been heavily explored for novel electronic phases(*8, 10–19*) and collective excitations(*20–25*). Current research on the latter is dominated by moiré excitons with distinct spatial geometries(*20–24*), which arise from valence-to-conduction-band transitions with energy scale of eV. By contrast, the low-lying electronic collective excitations arise from intra- and inter-miniband transitions(*26*) and govern low-energy properties of ground phases, which, however, remain largely unexplored experimentally, not to mention the CSM.

In this work, we report the observation of a gate-tunable CSM in near-0°-twisted $WSe_2/WS_2$ bilayers, by performing polarization-resolved Raman measurements. Specifically, by varying fillings (ν, the number of electrons per moiré unit cell), the CSM emerges as a sharp peak in the pseudovector-symmetry channel of the Raman spectra, arising from spin-flip transitions between conduction SOC-split minibands. Around $ν = 2$ that the sample is insulating, the mode is an electron-hole pair and hence chiral spin exciton, with frequency of ~13 cm$^{-1}$ and narrow linewidth of ~5 cm$^{-1}$. Notably, it has a giant Raman efficiency of about $1.2 \times 10^{-2}$ cm$^{-1}$Sr$^{-1}$, orders larger than phonon modes like G mode in graphene. As ν deviates from 2 that the sample is metallic, chiral spin excitonic polaron is formed by interacting with itinerant charges, exhibiting frequency increase up to 55 cm$^{-1}$ and intensity decrease. The spin-flip nature is further confirmed by the magnetic field dependence of the CSM that exhibits a Zeeman shift with a g-factor of about 0.34, which is consistent with the value of SOC-split conduction bands in TMD monolayers(*27*). In addition, the CSM vanishes for $ν \leq 1$, which compellingly evidences a charge-transfer(*28*) instead of Mott-Hubbard(*8*) insulator at $ν = 1$, providing a new method of clarifying the nature of insulating states for TMD moiré materials.

**The device and insulating states**
Fig. 1B shows the device schematic, a 0°-twisted $WSe_2/WS_2$ bilayer with a dual-gated geometry. There are three types of high-symmetry stacking sites, denoted by AA, $B^{W/Se}$, and $B^{S/W}$ (Fig. 1A). The bilayer belongs to the $C_{3v}$ point group and has a type-II band alignment(*24*) with electrons and holes respectively from $WS_2$ and $WSe_2$. Fig. 1D shows the doping-dependent reflectance contrast (*R*) spectrum near the resonance of the moiré intralayer excitons of $WSe_2$ for positive gate voltages (i.e., electron doping). Unless otherwise specified, we perform all optical measurements at a sample temperature of 3 K. The enhancement of $R$ is attributed to the formation of insulating states at specific fillings that are assigned to 1/3, 2/3, 1, 2, and 3 (Sect. 3 of SM and Fig. S1). In spite of intensive exploration of these insulating states(*10, 12, 14, 19, 29, 30*), it is still ambiguous whether the state at $\nu = 1$ is a Mott-Hubbard(*8*) or charge-transfer(*28*) insulator, due to the lack of experimental evidence for the charge distribution of $\nu > 1$. As explained below, our doping-dependent Raman measurements (Fig. 3) evidence a charge-transfer insulator at $\nu = 1$, with the charge distribution revealed by the density-function-theory(DFT) calculations, as depicted schematically in Fig. 1C—electrons successively occupying the $B^{W/Se}$ and $B^{S/W}$ sites.

**Probing the chiral spin mode**

Since being electric-dipole forbidden, the CSM cannot be probed by transmission or reflection spectroscopy. Instead, polarization-resolved Raman spectroscopy turns out to be a feasible choice(*31*). For example, in the $C_{3v}$ point group, there are three types of Raman modes with distinct symmetries classified by the irreducible representations(*32*), $A_1$, $A_2$, and E, which can be separated by varying the polarization geometry (Sect. 5 of SM). Since the basis functions of the $A_2$ representation transform as the out-of-plane projection of the angular momentum that is a pseudovector(*33*), the CSM with an out-of-plane spin moment will be associated with the $A_2$ channel(*5, 31*), and exhibits an antisymmetric Raman tensor. To be noted, the $A_2$-channel modes can also originate from transitions between crystal-field-split states with proper symmetries(*34–37*), which, however, are not relevant in the TMD materials based on symmetry consideration (Sect. 9 of SM). In addition, the modes associated with the $A_1$ and E channels have symmetric Raman tensors, such as breathing and shearing phonons(*38*).

We first show the polarization-resolved Raman spectra at $\nu = 2$ in Fig. 2A, with four polarization geometries, *XX*, *XY*, *RR* and *RL*, respectively representing parallel, perpendicular, co-circular, and cross-circular polarizations of the incident and scattered light. The excitation energy is 2.33 eV, close to the B exciton of $WS_2$ (the inset of Fig. 1D). The Stokes Raman scattering is shown, corresponding to a positive Raman shift. Strikingly, there is a strong and sharp resonance in the *XY* and *RR* geometries, in contrast to weak Raman signals in the *XX* and *RL* geometries. According to the $C_{3v}$ point group, the intensities can be separated into distinct symmetry channels as shown in Fig. 2B by: $I_{A_1} = I_{XX} - \frac{1}{2}I_{RL}$, $I_{A_2} = I_{XY} - \frac{1}{2}I_{RL}$, and $I_E = I_{RL}$ (Sect. 5 of SM). We can see that the strong resonance only exists in the $A_2$ channel, indicating a CSM. The $A_2$ representation is further confirmed by angle-resolved Raman spectroscopy (Fig. S3). This mode can be fit by a Voigt function with a smooth baseline (Sect. 6 of SM). Its frequency and deconvoluted full-width-half-maximum (FWHM) are about 13 and 5 cm$^{-1}$, respectively. In addition, the $A_1$ channel shows breathing phonons of the heterostructure(*39*) (< 50 cm$^{-1}$) and an out-of-plane vibration mode of Se atoms(*38*) (~ 250 cm$^{-1}$); the E channel shows negligible Raman signal.

**Chiral spin exciton and excitonic polaron**

To further understand the CSM, we need to explore its filling dependence, as shown in Fig. 3B (see all symmetry channels and extended filling range in Fig. S4). The CSM at varying fillings are analyzed by the Voigt fitting, from which the spectral weight and mode frequency are extracted (Fig. 3, C-E). The CSM can only be observed for electron doping with $\nu$ above 1 and below 3, with the strongest intensity and lowest frequency around $\nu$=2. As $\nu$ increases from 2, the frequency increases linearly up to 55 cm$^{-1}$ with the intensity monotonically decreasing; as $\nu$ decreases from 2, the frequency shows smaller increase up to 25 cm$^{-1}$, with an abrupt jump of about 5 cm$^{-1}$ around $\nu$=1.56, and the intensity generally decreases except for a local maximum around $\nu$=1.46. A similar filling dependence of the CSM has been reproduced in another near-0°-twisted WSe$_2$/WS$_2$ bilayer (Fig. S5). To be noted, no CSM has been observed in near-60°-twisted or angle-misaligned bilayers (Fig. S6 and Sect. 12 of SM).

The filling dependence in Fig. 3 evidences a charge-transfer(*28*) instead of Mott-Hubbard(*8*) insulator at $\nu = 1$. The strongest intensity at $\nu = 2$ and the emerging onset around $\nu = 1$ indicate that only the second electron, denoted by the sequence of filling a moiré unit cell, is directly involved in the CSM, i.e., no contributions from the first electron. Electronic Raman scattering is an orbital-sensitive technique(*40, 41*), as the light-matter interaction and Coulomb interaction in the scattering process highly depend on the orbitals involved. Therefore, the filling dependence evidences that the second electron has an orbit wavefunction distinct from the first electron, i.e., it is a charge-transfer insulator at $\nu = 1$. Otherwise, if it is a Mott-Hubbard insulator, despite the energy difference arising from the repulsive Coulomb interaction, the first and the second electrons would have the same orbital wavefunctions, and we should have observed the CSM at the beginning of doping, i.e., below filling of 1. Moreover, the charge-transfer insulator at $\nu = 1$ is consistent with the DFT calculation and energy estimation (Fig. S7 and Sect. 15 of SM), which also reveal the charge distribution schematically depicted in Fig. 1C. The first electron is localized at the B$^{W/Se}$ site based on the DFT calculation (Fig. S7), which is consistent with a previous scanning-tunneling microscope study(*30*); the second electron is mainly distributed over the B$^{S/W}$ site, owing to that the energy difference between the minibands from the B$^{S/W}$ and B$^{W/Se}$ sites is smaller than the on-site Coulomb energy at the B$^{W/Se}$ site (Sect. 15 of SM). We denote the lowest-energy minibands distributed over the B$^{W/Se}$ and B$^{S/W}$ sites as $C_1$ and $C_2$, respectively. Once the first and second electrons are filled, both minibands would split into the lower and upper many-body bands, labelled by the superscripts of $L$ and $U$, respectively.

The CSM could be further specified based on the filling—it is a chiral spin exciton at filling of 2, which transitions to a chiral spin exciton polaron at fillings away from 2. Specifically, as shown in Fig. 3A (top panel), for $\nu$=2 that the bilayer is insulating, the Raman scattering promotes one electron from the fully occupied $C_2^L$ band to the empty SOC-split band $C_{SO}$, leading to an electron-hole pair—*chiral spin exciton* (*X*). Since its energy (~13 cm$^{-1}$, about 1.6 meV) is much lower than the SOC-splitting of the conduction band of TMD [~30 meV, Ref. (*42, 43*)], the exciton has relatively large binding energy, which likely arises from the exchange Coulomb interaction (10s of meV) inherited from TMD monolayers(*43–45*). For $\nu \neq 2$, the exciton interacts with itinerant charges and forms *chiral spin excitonic polaron*, denoted by $X^-$ and $X^+$ for $\nu > 2$ and $\nu < 2$, respectively. The filling dependences of $X^-$ and $X^+$ arise from screening effects(*46, 47*) by itinerant charges, which is similar to the doping dependence of repulsive exciton polarons in TMD monolayers(*48*).

Furthermore, since the B exciton of $WS_2$ is close to the laser energy of 2.33 eV (the inset of Fig. 1D), we propose a B-exciton-involved mechanism for the Raman scattering by the CSM, concisely depicted in Fig. 3A (bottom panel). First, a B exciton is virtually created by the incident light; secondly, the B exciton is scattered by Coulomb interaction with the Fermi sea into another state, and meanwhile the CSM is created; finally, the B exciton recombines and emits the scattered light. The spin-flip process is enabled by the exchange part of Coulomb interaction(*41*). To be noted, we also observe the CSM upon another laser energy of 1.959 eV that is close to the A exciton of $WS_2$, for which we propose a similar A-exciton-involved mechanism. See details in Sect. 11 of SM and Fig. S8-10.

**The Zeeman effect of chiral spin mode**
To further confirm what we observe is indeed a CSM, we need to demonstrate its spin-flip nature. We study the $A_2$-channel Raman spectrum at $\nu=2$ upon an out-of-plane magnetic field, shown in Fig. 4A. The frequency of the CSM almost doesn't change below 2 T, but increases linearly from about 13 to 16 cm$^{-1}$ as the magnetic field increases from 2 to 9 T, revealing a g-factor of $0.34 \pm 0.02$ (Fig. 4B); concomitantly, the deconvoluted linewidth slightly broadens by about 1 cm$^{-1}$ (Fig. 4C). It should be noted that the $A_2$-channel Raman spectra at all magnetic fields are extracted from the *XY* geometry, in which the Raman scatterings occurring at the *K* and *K'* valleys are equally detected. We can also extract the CSM from other polarization geometries (e.g., *RR* and *LL*), which, however, only show slight intensity difference even at 9 T (Fig. S11), likely due to the lack of polarization-dependent valley selectivity of the CSM at excitation energy of 2.33 eV.

The magnetic field dependence of the CSM can be understood in terms of the spin Zeeman effect. Fig. 4D shows a schematic of the magnetic-field dependent band alignment at $\nu=2$. Upon a positive magnetic field, the $C_2^L$ and $C_{SO}$ bands at the *K(K')* valley move away from (towards) each other owing to the spin Zeeman effect, which leads to frequency increase(decrease) of the CSM. At small magnetic field, the electrons of the $C_2^L$ band are almost equally populated over the *K* and *K'* valleys. The CSM at two valleys cannot be separated in frequency due to the small splitting (~0.6 cm$^{-1}$ at 2 T using *g*-factor of 0.34) compared to the broad linewidth (~5 cm$^{-1}$), and hence the overall frequency changes little. At large magnetic field, the electrons of the $C_2^L$ band are *K*-valley polarized, and hence only one CSM peak from the *K* valley can be observed, with the frequency increasing linearly. The valley or orbital effects are not considered, as they will not cause any relative shift between the $C_2^L$ and $C_{SO}$ bands at the same valley(*27*). Notably, the g-factor of the CSM is $0.34 \pm 0.02$, very close to the theoretical g-factor of 0.38 for SOC-split conduction bands in TMD monolayers(*27*), strongly evidencing the spin-flip nature of the CSM.

**Temperature dependence and giant Raman efficiency**
To evaluate the thermal robustness of the CSM, we study $A_2$-channel Raman spectrum at $\nu=2$ upon varying temperatures, shown in Fig. 5A. As the temperature increases, the intensity of the CSM decreases to a half at ~25 K (Fig. 5C), which rules out the participation of phonons; concomitantly, the frequency increases to about 18 cm$^{-1}$ and the deconvoluted linewidth broadens by about 3 cm$^{-1}$ (Fig. 5, D, E). Meanwhile, we measure the filling dependent $R_{max}$ at varying temperatures (Fig. 5B), which reveal the thermal evolution of the insulating states. Specifically, the insulating state at $\nu=2$ that is characterized by the enhancement of $R_{max}$ (shaded area), shows a transition temperature of about 50 K (Fig. 5C), corresponding to a charge gap of ~20 meV(*49*), consistent with previous studies(*14, 19, 49*). The temperature dependences of the CSM and $R_{max}$ show quite similar behavior, suggesting that both of which can be attributed to screening effects by thermally excited carriers.

The relatively large binding energy of the chiral spin exciton at $\nu=2$ indicates strong light-matter interaction and hence high transition rate (*44*). Indeed, the Raman intensity of the CSM is surprisingly stronger than that of phonons in the heterobilayer (Fig. 2), especially considering the much lower density of doped electrons (~$10^{12}$ cm$^{-2}$) than that of atoms (~$10^{15}$ cm$^{-2}$). We further determine the Raman efficiency of the CSM by a sample-substitution method(*50*), in which we compare the Raman spectrum of the CSM with G phonon in graphene that has known Raman efficiency using the same experimental setup (Sect. 13 of SM and Fig. S12). The CSM shows a giant Raman efficiency of ~$1.2 \times 10^{-2}$ cm$^{-1}$Sr$^{-1}$, which is orders larger than many phonon modes, including G phonon (Table S1). Such giant Raman efficiency is likely associated with spatial overlap between quasiparticles in the presence of the moiré potential(*51*), e.g., between the B exciton of $WS_2$ and doped electrons, which leads to enhanced Coulomb interaction and hence efficient Raman scattering (Sect. 11 of SM).

**Conclusion**
In this work, we report the CSM in TMD moiré superlattices, which exhibits unprecedented tunability compared to previous studies(*5, 6*) and clarifies the nature of the insulating states(*8, 28*). The CSM has giant Raman efficiency and relatively large thermal robustness, which may be useful for gate-tunable Raman lasers(*52*) and novel magnonics(*6*). The small energy of the CSM is also promising for realizing triplet exciton insulators(*53*), through dielectric engineering(*54*) that can further enhance the binding energy. Moreover, our study should trigger a wide search for exotic low-lying collective excitations, especially in the systems with the interplay of the SOC and many-body interactions like twisted $MoTe_2$ bilayers(*11, 15–18*) and TMD-proximitized graphite(*55–58*), which will also shed light on the ground phases.

**References and Notes**

1. W. Witczak-Krempa, G. Chen, Y. B. Kim, L. Balents, Correlated Quantum Phenomena in the Strong Spin-Orbit Regime. *Annu. Rev. Condens. Matter Phys.* **5**, 57–82 (2014).
2. A. Ashrafi, D. L. Maslov, Chiral Spin Waves in Fermi Liquids with Spin-Orbit Coupling. *Phys. Rev. Lett.* **109**, 227201 (2012).
3. A. Shekhter, M. Khodas, A. M. Finkel'stein, Chiral spin resonance and spin-Hall conductivity in the presence of the electron-electron interactions. *Phys. Rev. B* **71**, 165329 (2005).
4. A. G. Gurevich, G. A. Melkov, *Magnetization Oscillations and Waves* (Taylor \& Francis, 1996).
5. H.-H. Kung, S. Maiti, X. Wang, S.-W. Cheong, D. L. Maslov, G. Blumberg, Chiral Spin Mode on the Surface of a Topological Insulator. *Phys. Rev. Lett.* **119**, 136802 (2017).
6. F. Perez, F. Baboux, C. A. Ullrich, I. D'Amico, G. Vignale, G. Karczewski, T. Wojtowicz, Spin-Orbit Twisted Spin Waves: Group Velocity Control. *Phys. Rev. Lett.* **117**, 137204 (2016).
7. K. F. Mak, J. Shan, Semiconductor moiré materials. *Nat. Nanotechnol.* **17**, 686–695 (2022).
8. F. Wu, T. Lovorn, E. Tutuc, A. H. MacDonald, Hubbard model physics in transition metal dichalcogenide moiré bands. *Phys. Rev. Lett.* **121**, 26402 (2018).
9. D. Xiao, G. Bin Liu, W. Feng, X. Xu, W. Yao, Coupled spin and valley physics in monolayers of $MoS_2$ and other group-VI dichalcogenides. *Phys. Rev. Lett.* **108**, 1–5 (2012).
10. E. C. Regan, D. Wang, C. Jin, M. I. Bakti Utama, B. Gao, X. Wei, S. Zhao, W. Zhao, Z. Zhang, K. Yumigeta, M. Blei, J. D. Carlström, K. Watanabe, T. Taniguchi, S. Tongay, M. Crommie, A. Zettl, F. Wang, Mott and generalized Wigner crystal states in $WSe_2/WS_2$ moiré superlattices. *Nature* **579**, 359–363 (2020).
11. F. Wu, T. Lovorn, E. Tutuc, I. Martin, A. H. Macdonald, Topological Insulators in Twisted Transition Metal Dichalcogenide Homobilayers. *Phys. Rev. Lett.* **122**, 86402 (2019).
12. Y. Tang, L. Li, T. Li, Y. Xu, S. Liu, K. Barmak, K. Watanabe, T. Taniguchi, A. H. MacDonald, J. Shan, K. F. Mak, Simulation of Hubbard model physics in $WSe_2/WS_2$ moiré superlattices. *Nature* **579**, 353–358 (2020).
13. T. Li, S. Jiang, B. Shen, Y. Zhang, L. Li, Z. Tao, T. Devakul, K. Watanabe, T. Taniguchi, L. Fu, J. Shan, K. F. Mak, Quantum anomalous Hall effect from intertwined moiré bands. *Nature* **600**, 641–646 (2021).
14. Y. Xu, S. Liu, D. A. Rhodes, K. Watanabe, T. Taniguchi, J. Hone, V. Elser, K. F. Mak, J. Shan, Correlated insulating states at fractional fillings of moiré superlattices. *Nature* **587**, 214–218 (2020).
15. J. Cai, E. Anderson, C. Wang, X. Zhang, X. Liu, W. Holtzmann, Y. Zhang, F. Fan, T. Taniguchi, K. Watanabe, Y. Ran, T. Cao, L. Fu, D. Xiao, W. Yao, X. Xu, Signatures of fractional quantum anomalous Hall states in twisted $MoTe_2$. *Nature* **622**, 63–68 (2023).

16. Y. Zeng, Z. Xia, K. Kang, J. Zhu, P. Knüppel, C. Vaswani, K. Watanabe, T. Taniguchi, K. F. Mak, J. Shan, Thermodynamic evidence of fractional Chern insulator in moiré $MoTe_2$. *Nature* **622**, 69–73 (2023).
17. H. Park, J. Cai, E. Anderson, Y. Zhang, J. Zhu, X. Liu, C. Wang, W. Holtzmann, C. Hu, Z. Liu, T. Taniguchi, K. Watanabe, J.-H. Chu, T. Cao, L. Fu, W. Yao, C.-Z. Chang, D. Cobden, D. Xiao, X. Xu, Observation of fractionally quantized anomalous Hall effect. *Nature* **622**, 74–79 (2023).
18. F. Xu, Z. Sun, T. Jia, C. Liu, C. Xu, C. Li, Y. Gu, K. Watanabe, T. Taniguchi, B. Tong, J. Jia, Z. Shi, S. Jiang, Y. Zhang, X. Liu, T. Li, Observation of Integer and Fractional Quantum Anomalous Hall Effects in Twisted Bilayer $MoTe_2$. *Phys. Rev. X* **13**, 31037 (2023).
19. X. Huang, T. Wang, S. Miao, C. Wang, Z. Li, Z. Lian, T. Taniguchi, K. Watanabe, S. Okamoto, D. Xiao, S.-F. Shi, Y.-T. Cui, Correlated insulating states at fractional fillings of the $WS_2/WSe_2$ moiré lattice. *Nat. Phys.* **17**, 715–719 (2021).
20. C. Jin, E. C. Regan, A. Yan, M. Iqbal Bakti Utama, D. Wang, S. Zhao, Y. Qin, S. Yang, Z. Zheng, S. Shi, K. Watanabe, T. Taniguchi, S. Tongay, A. Zettl, F. Wang, Observation of moiré excitons in $WSe_2/WS_2$ heterostructure superlattices. *Nature* **567**, 76–80 (2019).
21. E. M. Alexeev, D. A. Ruiz-Tijerina, M. Danovich, M. J. Hamer, D. J. Terry, P. K. Nayak, S. Ahn, S. Pak, J. Lee, J. I. Sohn, M. R. Molas, M. Koperski, K. Watanabe, T. Taniguchi, K. S. Novoselov, R. V Gorbachev, H. S. Shin, V. I. Fal'ko, A. I. Tartakovskii, Resonantly hybridized excitons in moiré superlattices in van der Waals heterostructures. *Nature* **567**, 81–86 (2019).
22. K. Tran, G. Moody, F. Wu, X. Lu, J. Choi, K. Kim, A. Rai, D. A. Sanchez, J. Quan, A. Singh, J. Embley, A. Zepeda, M. Campbell, T. Autry, T. Taniguchi, K. Watanabe, N. Lu, S. K. Banerjee, K. L. Silverman, S. Kim, E. Tutuc, L. Yang, A. H. MacDonald, X. Li, Evidence for moiré excitons in van der Waals heterostructures. *Nature* **567**, 71–75 (2019).
23. K. L. Seyler, P. Rivera, H. Yu, N. P. Wilson, E. L. Ray, D. G. Mandrus, J. Yan, W. Yao, X. Xu, Signatures of moiré-trapped valley excitons in $MoSe_2/WSe_2$ heterobilayers. *Nature* **567**, 66–70 (2019).
24. Y. Tang, J. Gu, S. Liu, K. Watanabe, T. Taniguchi, J. Hone, K. F. Mak, J. Shan, Tuning layer-hybridized moiré excitons by the quantum-confined Stark effect. *Nat. Nanotechnol.* **16**, 52–57 (2021).
25. F. Wu, T. Lovorn, A. H. MacDonald, Topological Exciton Bands in Moiré Heterojunctions. *Phys. Rev. Lett.* **118**, 147401 (2017).
26. N. Saigal, L. Klebl, H. Lambers, S. Bahmanyar, V. Antić, D. M. Kennes, T. O. Wehling, U. Wurstbauer, Collective Charge Excitations between Moiré Minibands in Twisted $WSe_2$ Bilayers Probed with Resonant Inelastic Light Scattering. *Phys. Rev. Lett.* **133**, 46902 (2024).
27. M. Koperski, M. R. Molas, A. Arora, K. Nogajewski, M. Bartos, J. Wyzula, D. Vaclavkova, P. Kossacki, M. Potemski, Orbital, spin and valley contributions to Zeeman splitting of excitonic resonances in $MoSe_2$, $WSe_2$ and $WS_2$ Monolayers.

28. Y. Zhang, N. F. Q. Yuan, L. Fu, Moiré quantum chemistry: Charge transfer in transition metal dichalcogenide superlattices. *Phys. Rev. B* **102**, 201115 (2020).
29. C. Jin, Z. Tao, T. Li, Y. Xu, Y. Tang, J. Zhu, S. Liu, K. Watanabe, T. Taniguchi, J. C. Hone, L. Fu, J. Shan, K. F. Mak, Stripe phases in $WSe_2/WS_2$ moiré superlattices. *Nat. Mater.* **20**, 940–944 (2021).
30. H. Li, S. Li, E. C. Regan, D. Wang, W. Zhao, S. Kahn, K. Yumigeta, M. Blei, T. Taniguchi, K. Watanabe, S. Tongay, A. Zettl, M. F. Crommie, F. Wang, Imaging two-dimensional generalized Wigner crystals. *Nature* **597**, 650–654 (2021).
31. S. Maiti, D. L. Maslov, Raman scattering in a two-dimensional Fermi liquid with spin-orbit coupling. *Phys. Rev. B* **95**, 134425 (2017).
32. W. Hayes, R. Loudon, *Scattering of Light by Crystals* (Wiley, 1978).
33. M. S. Dresselhaus, G. Dresselhaus, A. Jorio, *Group Theory: Application to the Physics of Condensed Matter* (Springer, 2007).
34. Y. Wang, I. Petrides, G. McNamara, M. M. Hosen, S. Lei, Y.-C. Wu, J. L. Hart, H. Lv, J. Yan, D. Xiao, J. J. Cha, P. Narang, L. M. Schoop, K. S. Burch, Axial Higgs mode detected by quantum pathway interference in $RTe_3$. *Nature* **606**, 896–901 (2022).
35. H.-H. Kung, R. E. Baumbach, E. D. Bauer, V. K. Thorsmølle, W.-L. Zhang, K. Haule, J. A. Mydosh, G. Blumberg, Chirality density wave of the "hidden order" phase in $URu_2Si_2$. *Science* **347**, 1339–1342 (2015).
36. J. Buhot, M.-A. Méasson, Y. Gallais, M. Cazayous, A. Sacuto, G. Lapertot, D. Aoki, Symmetry of the Excitations in the Hidden Order State of $URu_2Si_2$. *Phys. Rev. Lett.* **113**, 266405 (2014).
37. H. Rho, M. V Klein, P. C. Canfield, Polarized Raman scattering studies of crystal-field excitations in $ErNi_2B_2C$. *Phys. Rev. B* **69**, 144420 (2004).
38. Y. Zhao, X. Luo, H. Li, J. Zhang, P. T. Araujo, C. K. Gan, J. Wu, H. Zhang, S. Y. Quek, M. S. Dresselhaus, Q. Xiong, Interlayer Breathing and Shear Modes in Few-Trilayer $MoS_2$ and $WSe_2$. *Nano Lett.* **13**, 1007–1015 (2013).
39. M.-L. Lin, Y. Zhou, J.-B. Wu, X. Cong, X.-L. Liu, J. Zhang, H. Li, W. Yao, P.-H. Tan, Cross-dimensional electron-phonon coupling in van der Waals heterostructures. *Nat. Commun.* **10**, 2419 (2019).
40. T. P. Devereaux, R. Hackl, Inelastic light scattering from correlated electrons. *Rev. Mod. Phys.* **79**, 175–233 (2007).
41. C. Schüller, *Inelastic Light Scattering of Semiconductor Nanostructures: Fundamentals and Recent Advances* (Springer Berlin Heidelberg, 2006).
42. G.-B. Liu, W.-Y. Shan, Y. Yao, W. Yao, D. Xiao, Three-band tight-binding model for monolayers of group-VIB transition metal dichalcogenides. *Phys. Rev. B* **88**, 85433 (2013).
43. J. P. Echeverry, B. Urbaszek, T. Amand, X. Marie, I. C. Gerber, Splitting between bright and dark excitons in transition metal dichalcogenide monolayers. *Phys. Rev. B* **93**, 121107 (2016).
44. G. Wang, A. Chernikov, M. M. Glazov, T. F. Heinz, X. Marie, T. Amand, B. Urbaszek, Colloquium: Excitons in atomically thin transition metal dichalcogenides. *Rev. Mod. Phys.* **90**, 21001 (2018).

45. M. Zinkiewicz, A. O. Slobodeniuk, T. Kazimierczuk, P. Kapuściński, K. Oreszczuk, M. Grzeszczyk, M. Bartos, K. Nogajewski, K. Watanabe, T. Taniguchi, C. Faugeras, P. Kossacki, M. Potemski, A. Babiński, M. R. Molas, Neutral and charged dark excitons in monolayer $WS_2$. *Nanoscale* **12**, 18153–18159 (2020).

46. C. Fey, P. Schmelcher, A. Imamoglu, R. Schmidt, Theory of exciton-electron scattering in atomically thin semiconductors. *Phys. Rev. B* **101**, 195417 (2020).

47. D. Van Tuan, B. Scharf, I. Žutić, H. Dery, Marrying Excitons and Plasmons in Monolayer Transition-Metal Dichalcogenides. *Phys. Rev. X* **7**, 41040 (2017).

48. D. Huang, K. Sampson, Y. Ni, Z. Liu, D. Liang, K. Watanabe, T. Taniguchi, H. Li, E. Martin, J. Levinsen, M. M. Parish, E. Tutuc, D. K. Efimkin, X. Li, Quantum Dynamics of Attractive and Repulsive Polarons in a Doped $MoSe_2$ Monolayer. *Phys. Rev. X* **13**, 11029 (2023).

49. X. Huang, D. Chen, Z. Lian, Q. Wu, M. Rashetnia, M. Blei, T. Taniguchi, K. Watanabe, S. Tongay, S.-F. Shi, Y.-T. Cui, Measurements of Correlated Insulator Gaps in a Transition Metal Dichalcogenide Moiré Superlattice. *arXiv:2308.08134* (2023).

50. P. Klar, E. Lidorikis, A. Eckmann, I. A. Verzhbitskiy, A. C. Ferrari, C. Casiraghi, Raman scattering efficiency of graphene. *Phys. Rev. B* **87**, 205435 (2013).

51. B. Evrard, H. S. Adlong, A. A. Ghita, T. Uto, L. Ciorciaro, K. Watanabe, T. Taniguchi, M. Kroner, A. İmamoğlu, Nonlinear spectroscopy of semiconductor moiré materials. *arXiv:2402.16630* (2024).

52. C. K. N. Patel, E. D. Shaw, R. J. Kerl, Tunable Spin-Flip Laser and Infrared Spectroscopy. *Phys. Rev. Lett.* **25**, 8–11 (1970).

53. Z. Jiang, W. Lou, Y. Liu, Y. Li, H. Song, K. Chang, W. Duan, S. Zhang, Spin-Triplet Excitonic Insulator: The Case of Semihydrogenated Graphene. *Phys. Rev. Lett.* **124**, 166401 (2020).

54. A. Chaves, J. G. Azadani, H. Alsalman, D. R. da Costa, R. Frisenda, A. J. Chaves, S. H. Song, Y. D. Kim, D. He, J. Zhou, A. Castellanos-Gomez, F. M. Peeters, Z. Liu, C. L. Hinkle, S.-H. Oh, P. D. Ye, S. J. Koester, Y. H. Lee, P. Avouris, X. Wang, T. Low, Bandgap engineering of two-dimensional semiconductor materials. *npj 2D Mater. Appl.* **4**, 29 (2020).

55. Y. Sha, J. Zheng, K. Liu, H. Du, K. Watanabe, T. Taniguchi, J. Jia, Z. Shi, R. Zhong, G. Chen, Observation of a Chern insulator in crystalline ABCA-tetralayer graphene with spin-orbit coupling. *Science* **384**, 414–419 (2024).

56. H. S. Arora, R. Polski, Y. Zhang, A. Thomson, Y. Choi, H. Kim, Z. Lin, I. Z. Wilson, X. Xu, J.-H. Chu, K. Watanabe, T. Taniguchi, J. Alicea, S. Nadj-Perge, Superconductivity in metallic twisted bilayer graphene stabilized by WSe2. *Nature* **583**, 379–384 (2020).

57. T. Han, Z. Lu, Y. Yao, J. Yang, J. Seo, C. Yoon, K. Watanabe, T. Taniguchi, L. Fu, F. Zhang, L. Ju, Large quantum anomalous Hall effect in spin-orbit proximitized rhombohedral graphene. *Science* **384**, 647–651 (2024).

58. Y. Zhang, R. Polski, A. Thomson, É. Lantagne-Hurtubise, C. Lewandowski, H.

Zhou, K. Watanabe, T. Taniguchi, J. Alicea, S. Nadj-Perge, Enhanced superconductivity in spin–orbit proximitized bilayer graphene. *Nature* **613**, 268–273 (2023).

59. L. Wang, I. Meric, P. Y. Huang, Q. Gao, Y. Gao, H. Tran, T. Taniguchi, K. Watanabe, L. M. Campos, D. A. Muller, J. Guo, P. Kim, J. Hone, K. L. Shepard, C. R. Dean, One-dimensional electrical contact to a two-dimensional material. *Science* **342**, 614–617 (2013).
60. X. Wang, X. Zhang, J. Zhu, H. Park, Y. Wang, C. Wang, W. G. Holtzmann, T. Taniguchi, K. Watanabe, J. Yan, D. R. Gamelin, W. Yao, D. Xiao, T. Cao, X. Xu, Intercell moiré exciton complexes in electron lattices. *Nat. Mater.* **22**, 599–604 (2023).
61. Z. Lian, Y. Meng, L. Ma, I. Maity, L. Yan, Q. Wu, X. Huang, D. Chen, X. Chen, X. Chen, M. Blei, T. Taniguchi, K. Watanabe, S. Tongay, J. Lischner, Y.-T. Cui, S.-F. Shi, Valley-polarized excitonic Mott insulator in WS2/WSe2 moiré superlattice. *Nat. Phys.* **20**, 34–39 (2024).
62. S. D. M. Brown, A. Jorio, P. Corio, M. S. Dresselhaus, G. Dresselhaus, R. Saito, K. Kneipp, Origin of the Breit-Wigner-Fano lineshape of the tangential G-band feature of metallic carbon nanotubes. *Phys. Rev. B* **63**, 155414 (2001).
63. M. Cardona, G. Guntherodt, "Light Scattering in Solids IV: Electronic Scattering, Spin Effects, SERS, and Morphic Effects" (Springer, 1984), pp. 5–150.
64. A. Kormányos, G. Burkard, M. Gmitra, J. Fabian, V. Zólyomi, N. D. Drummond, V. Fal'ko, k·p theory for two-dimensional transition metal dichalcogenide semiconductors. *2D Mater.* **2**, 022001 (2015).
65. K. Ishito, H. Mao, K. Kobayashi, Y. Kousaka, Y. Togawa, H. Kusunose, J. Kishine, T. Satoh, Chiral phonons: circularly polarized Raman spectroscopy and ab initio calculations in a chiral crystal tellurium. *Chirality* **35**, 338–345 (2023).
66. D. G. Thomas, J. J. Hopfield, Spin-Flip Raman Scattering in Cadmium Sulfide. *Phys. Rev.* **175**, 1021–1032 (1968).
67. M. H. Naik, E. C. Regan, Z. Zhang, Y.-H. Chan, Z. Li, D. Wang, Y. Yoon, C. S. Ong, W. Zhao, S. Zhao, M. I. B. Utama, B. Gao, X. Wei, M. Sayyad, K. Yumigeta, K. Watanabe, T. Taniguchi, S. Tongay, F. H. da Jornada, F. Wang, S. G. Louie, Intralayer charge-transfer moiré excitons in van der Waals superlattices. *Nature* **609**, 52–57 (2022).
68. H. C. P. Movva, B. Fallahazad, K. Kim, S. Larentis, T. Taniguchi, K. Watanabe, S. K. Banerjee, E. Tutuc, Density-Dependent Quantum Hall States and Zeeman Splitting in Monolayer and Bilayer WSe$_2$. *Phys. Rev. Lett.* **118**, 247701 (2017).
69. Y. Li, A. Chernikov, X. Zhang, A. Rigosi, H. M. Hill, A. M. van der Zande, D. A. Chenet, E.-M. Shih, J. Hone, T. F. Heinz, Measurement of the optical dielectric function of monolayer transition-metal dichalcogenides: MoS$_2$, MoSe$_2$, WS$_2$, and WSe$_2$. *Phys. Rev. B* **90**, 205422 (2014).
70. I. H. Malitson, Refraction and Dispersion of Synthetic Sapphire. *J. Opt. Soc. Am.* **52**, 1377–1379 (1962).
71. V. Michaud-Rioux, L. Zhang, H. Guo, RESCU: A real space electronic structure method. *J. Comput. Phys.* **307**, 593–613 (2016).

72. H. Li, S. Li, M. H. Naik, J. Xie, X. Li, J. Wang, E. Regan, D. Wang, W. Zhao, S. Zhao, S. Kahn, K. Yumigeta, M. Blei, T. Taniguchi, K. Watanabe, S. Tongay, A. Zettl, S. G. Louie, F. Wang, M. F. Crommie, Imaging moiré flat bands in three-dimensional reconstructed $WSe_2/WS_2$ superlattices. *Nat. Mater.* **20**, 945–950 (2021).
73. J. Perdew, K. Burke, M. Ernzerhof, Generalized Gradient Approximation Made Simple. *Phys. Rev. Lett.* **77**, 3865–3868 (1996).
74. K. Momma, F. Izumi, VESTA 3 for three-dimensional visualization of crystal, volumetric and morphology data. *J. Appl. Crystallogr.* **44**, 1272–1276 (2011).
75. H. Pan, F. Wu, S. Das Sarma, Band topology, Hubbard model, Heisenberg model, and Dzyaloshinskii-Moriya interaction in twisted bilayer $WSe_2$. *Phys. Rev. Res.* **2**, 33087 (2020).
76. M. Cardona, M. H. Brodsky, *Light Scattering in Solids: Introductory Concepts* (Springer-Verlag, 1983).

**Acknowledgement:** We thank John A. McGuire for critical reading, Shaowei Li, Liguo Ma, Yunhao Lu for helpful discussions. **Funding:** This work was supported by the National Key R&D Program of China (grant nos. 2022YFA1402400 and 2022YFA1405400), the National Natural Science Foundation of China (grant nos. 12274365 and 12374163), Zhejiang Provincial Natural Science Foundation of China (grant no. LR24A040001), and Open project of Key Laboratory of Artificial Structures and Quantum Control (Ministry of Education) of Shanghai Jiao Tong University. K.W. and T.T. acknowledge support from the JSPS KAKENHI (grant nos. 20H00354 and 23H02052) and World Premier International Research Center Initiative (WPI), MEXT, Japan. In addition, we appreciate the device fabrication support from the ZJU Micro-Nano Fabrication Center in Zhejiang University.

**Author contributions:** Y.T. designed the scientific objectives. Y.T. and S.Z. supervised the project. Z.S. and W.S. fabricated the devices, and performed optical measurements. Y.X. performed DFT calculations. K.W. and T.T. grew the hBN crystals. Y.T. and Z.S. wrote the manuscript with the input from all the authors.

**Competing interests:** The authors declare no competing interests.

**Supplementary Materials**

Materials and Methods

Tables S1 to S2

Figs. S1 to S17

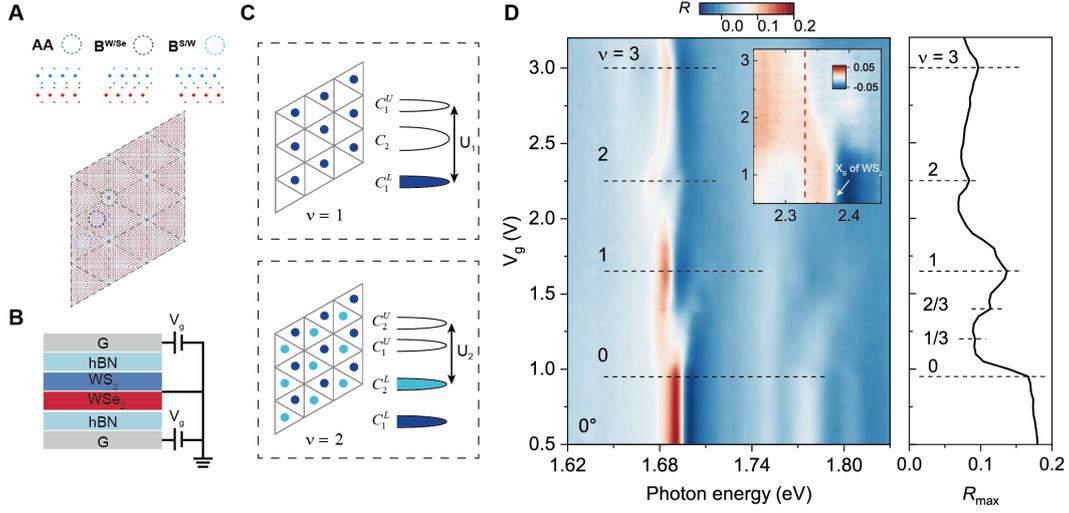

**Fig. 1 | The device schematic and insulating states in near-0°-twisted WSe₂/WS₂ moiré superlattices. A**, Schematic of moiré superlattices that have three high-symmetry stacking sites, denoted as AA, $B^{W/Se}$, and $B^{S/W}$. **B**, Schematic of a dual-gated device. The WSe₂/WS₂ bilayer is grounded, and a gate voltage $V_g$ is applied to the top and bottom gate. **C**, Schematic of the charge distribution. For $\nu=1$, electrons only reside at $B^{W/Se}$ sites; for $\nu=2$, additional electrons reside at the $B^{S/W}$ sites. $C_1$ and $C_2$ denote the lowest-energy minibands emerging from the $B^{W/Se}$ and $B^{S/W}$ sites, respectively. $C_{1(2)}^L$ and $C_{1(2)}^U$ denote the lower and upper many-body bands of $C_{1(2)}$, respectively.

**D**, The left panel is a contour plot of doping dependence of reflectance contrast ($R$) spectrum, from which the maximum of $R$, $R_{max}$, is extracted within the photon energy range from 1.65 to 1.72 eV (right panel). The inset shows the $R$ spectrum around the laser energy of 2.33 eV (dashed) used in Raman measurements, which is close to the B exciton of WS₂ ($X_B$). Horizontal dashed lines denote representative fillings.

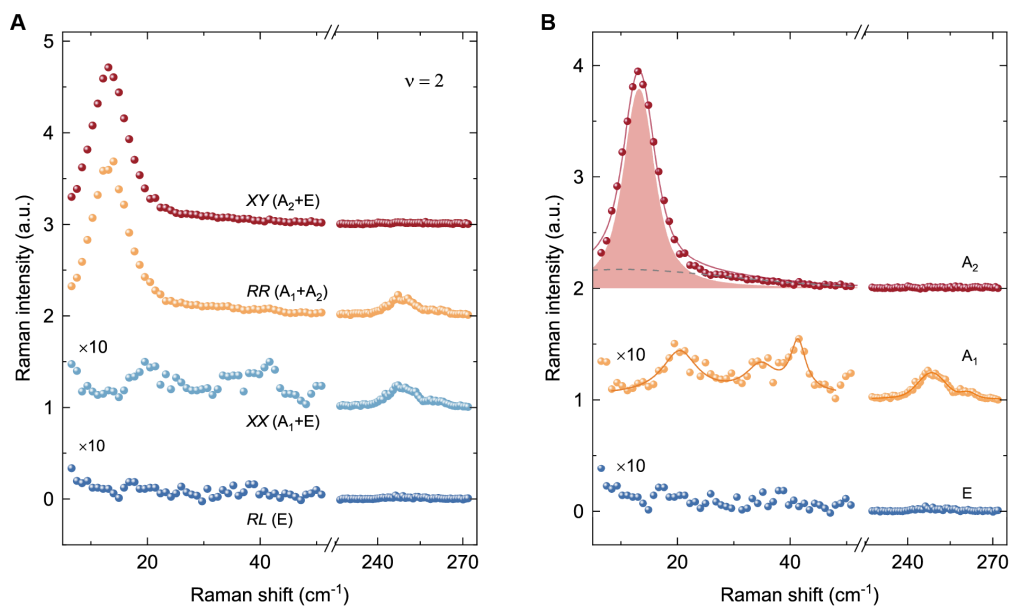

**Fig. 2 | The chiral spin mode at $\nu=2$ revealed by polarization-resolved Raman spectroscopy. A,** The Raman spectra at $\nu=2$ with four polarization geometries—*XX*, *XY*, *RR* and *RL*. Each polarization geometry contains given Raman modes, indicated by the formulas in the parentheses. **B,** Raman spectra in the $A_2$, $A_1$ and E channels, extracted from **A**. The $A_2$ channel shows intense Raman signal with frequency of about 13 cm$^{-1}$, fitted by a Voigt function (shaded), along with a smooth baseline (dashed). The spectra are vertically displaced for clarity. The scaling factors are only applied to the spectra below 50 cm$^{-1}$. The instrumental resolution is about 3.5 cm$^{-1}$.

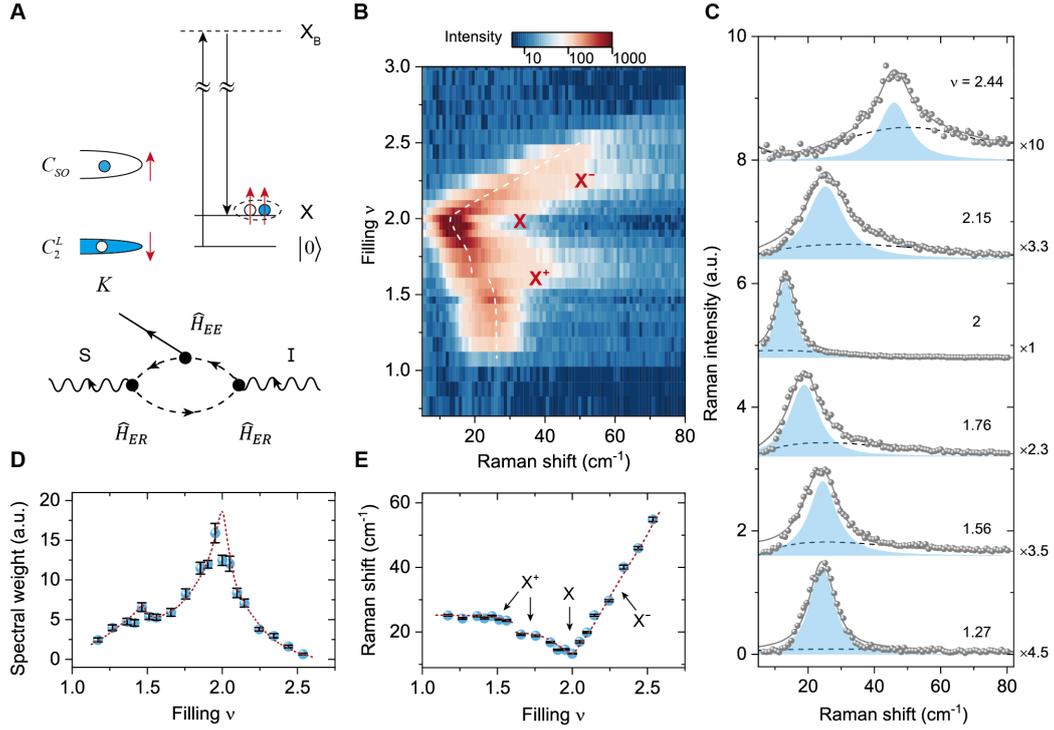

**Fig. 3 | Chiral spin exciton and excitonic polaron. A,** Schematic of the Raman excitation of the chiral spin exciton (X) at $\nu=2$. The exciton is composed of one electron and hole respectively from the SOC-split band $C_{SO}$ and $C_2^L$, which is generated by the Stokes Raman scattering via an intermediate state—the B exciton of $WS_2$ ($X_B$). Only the $K$ valley is shown for simplicity. Red arrows indicate the spin alignments. The diagrammatic representation (lower panel) illustrates the scattering mechanism, where I and S denote the incident and scattered photon, respectively, $\hat{H}_{ER}$ denotes the electron-radiation interaction, and $\hat{H}_{EE}$ denotes the Coulomb interaction between the exciton and the Fermi sea. **B,** The contour plot of the $A_2$-channel Raman spectrum on a log scale as a function of fillings. $X^-$ and $X^+$ denote chiral spin excitonic polaron at $\nu>2$ and $\nu<2$, respectively. **C,** The line cuts of **B** at representative fillings. The spectra are rescaled and vertically displaced for clarity. All spectra are fit by a Voigt function (blue shaded) with a smooth baseline (dashed line), indicated by solid lines. **D, E,** The spectral weight and frequency of the CSM as a function of fillings. Error bars are from the Voigt fitting. The white and red dashed curves are eye-guidance.

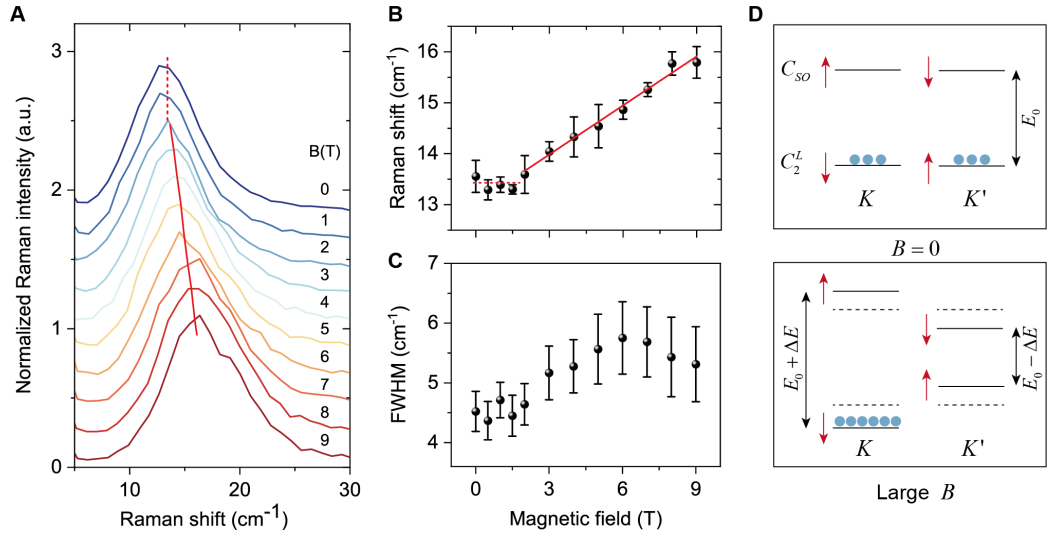

**Fig. 4 | The Zeeman effect of chiral spin mode. A**, The normalized $A_2$-channel Raman spectra at $\nu=2$ upon varying magnetic fields. Multiple scans of the magnetic field dependence are performed, and one representative spectrum is shown for each magnetic field (**Sect. 4** of **SM**). The spectra are vertically displaced. **B, C**, The frequency and deconvoluted FWHM of the CSM as a function of magnetic field. The error bars represent statistical errors from multiple scans. The horizontal dashed line is eye-guidance to the data at small magnetic fields (<2 T), indicating almost no frequency shift; the solid line is a linear fit to the data at large magnetic fields (≥2 T), revealing a g-factor of 0.34±0.02. **D**, Schematic of the magnetic field modulation of the band alignment and population at the $K$ and $K'$ valleys, only considering spin Zeeman effect. See main text for details.

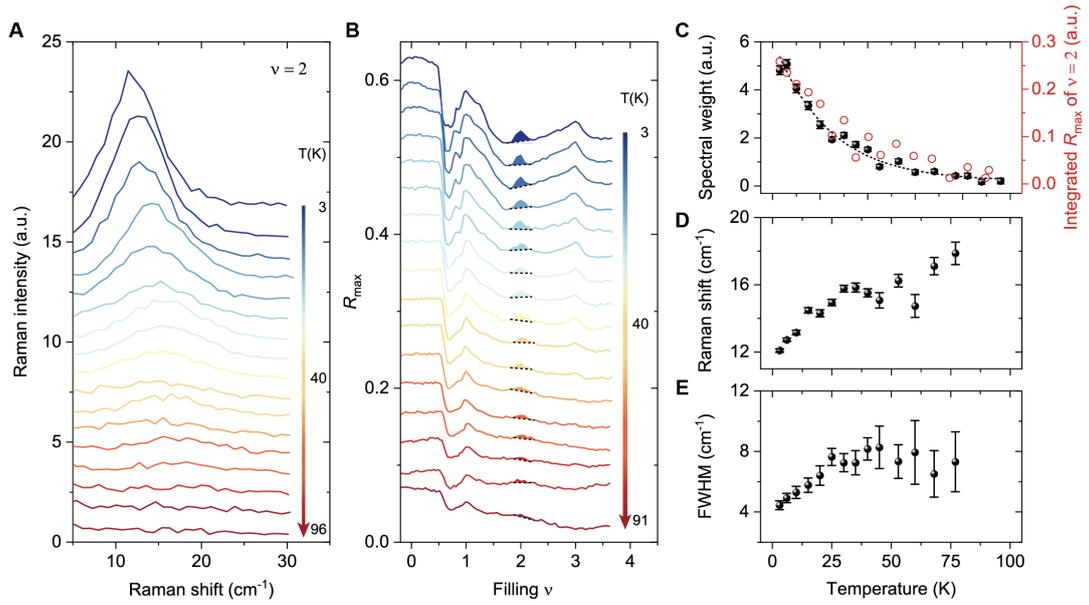

**Fig. 5 | The temperature dependences of chiral spin mode and the insulating state at $\nu=2$. A, B**, The $A_2$-channel Raman spectrum at $\nu=2$ and filling dependent $R_{max}$ upon varying temperatures. All curves are vertically displaced for clarity. The shaded denotes enhancement of $R_{max}$, arising from the insulating state at $\nu=2$. **C-E**, The temperature dependences of the spectral weight (**C**), frequency (**D**) and deconvoluted FWHM (**E**) of the CSM, denoted by black solid spheres. The dashed line is eye guidance to the integrated Raman intensity. The temperature dependence of the integral of $R_{max}$ around $\nu=2$, denoted by red empty circles, is overlaid with the Raman intensity. Error bars are from the Voigt fitting.